\documentclass[preprintnumbers,superscriptaddress,twocolumn,showpacs,preprintnumbers,amsmath,amssymb,prb]{revtex4}

\usepackage{graphicx}
\usepackage{dcolumn}
\usepackage{bm}
\usepackage{amsmath}

\newcommand{\ii}{\ensuremath{i}}
\newcommand{\dd}{\ensuremath{d}}
\begin{document}


\title{Noise-induced spectral shift measured in a double quantum dot}

\author{B. K\"ung}
\email{kuengb@phys.ethz.ch}
\affiliation{Solid State Physics Laboratory, ETH Zurich, 8093 Zurich, Switzerland}
\author{S. Gustavsson}
\affiliation{Solid State Physics Laboratory, ETH Zurich, 8093 Zurich, Switzerland}
\author{T. Choi}
\affiliation{Solid State Physics Laboratory, ETH Zurich, 8093 Zurich, Switzerland}
\author{I. Shorubalko}
\affiliation{Solid State Physics Laboratory, ETH Zurich, 8093 Zurich, Switzerland}
\affiliation{Electronics/Metrology/Reliability Laboratory, EMPA, 8600 D\"ubendorf, Switzerland}
\author{T. Ihn}
\affiliation {Solid State Physics Laboratory, ETH Zurich, 8093 Zurich, Switzerland}
\author{S. Sch\"on}
\affiliation {FIRST Laboratory, ETH Zurich, 8093 Zurich, Switzerland}
\author{F. Hassler}
\affiliation {Theoretische Physik, ETH Zurich, 8093 Zurich, Switzerland}
\author{G. Blatter}
\affiliation {Theoretische Physik, ETH Zurich, 8093 Zurich, Switzerland}
\author{K. Ensslin}
\affiliation {Solid State Physics Laboratory, ETH Zurich, 8093 Zurich, Switzerland}

\date{September 21, 2009}

\begin{abstract}
We measure the shot noise of a quantum point-contact using a capacitively coupled InAs double quantum dot as an on-chip
sensor. Our measurement signals are the (bidirectional) interdot electronic tunneling rates which are determined by
means of time-resolved charge sensing. The detector frequency is set by the relative detuning of the energy levels in
the two dots. For nonzero detuning, the noise in the quantum point-contact generates inelastic tunneling in the double
dot and thus causes an increase in the interdot tunneling rate. Conservation of spectral weight in the dots implies
that this increase must be compensated by a decrease in the rate close to zero detuning, which is quantitatively
confirmed in our experiment.
\end{abstract}

\pacs{73.63.Kv, 73.63.Nm, 72.70.+m, 73.23.Hk}

\maketitle

Charge detection with on-chip sensors provides a powerful tool for investigating the electronic properties of
mesoscopic circuits. By performing the detection with sufficient bandwidth, the observation of single-electron charging
events in real time becomes possible, which has been used, e.g., to read out the spin of quantum dots,\cite{Elzerman04}
to investigate the transport statistics of interacting electrons,\cite{Gustavsson06a, Fujisawa06} or to measure small
currents.\cite{Bylander05, Fujisawa06, Gustavsson08a} One of the simplest detectors offering enough sensitivity to
perform this kind of experiments is the quantum point-contact\cite{Field93} (QPC). However, the quantum dots (QDs) that
are typically probed by QPC sensors represent highly sensitive electronic devices on their own. Charge detection
therefore comes with a considerable amount of back-action of the QPC on the QD to which both
photons\cite{Gustavsson07a, Gustavsson08b} and phonons\cite{Khrapai06, Gasser09} have been shown to contribute.

One way of describing the photonic part of the back-action is in terms of the shot noise of the QPC which couples
capacitively to the QD system and generates photon-assisted tunneling\cite{Aguado00} (PAT). From this viewpoint, the QD
system can serve as a \emph{measurement} device for the QPC noise.\cite{Onac06a}$^,$\cite{Gustavsson08a} Since it works
on chip, it is inherently fast, and when using a double quantum dot (DQD), frequency-tunable noise detection becomes
possible via control of the interdot level detuning. In the work presented here, we use such a DQD detector to measure
noise of a QPC. Owing to the sample design, with the QPC located in a different host crystal (GaAs/AlGaAs) than the
noise probe (an InAs DQD), our setup features a suppression of the phononic part of the QPC-DQD interaction while
maintaining an extraordinarily large capacitive coupling: QPC conductance changes caused by dot charging can exceed
50\%, while the corresponding figure for split-gate or AFM-defined samples is typically a few
percent.\cite{Vandersypen04,Schleser04} In contrast to previous experiments, we are able to measure the response of the
DQD along the whole detuning axis from positive to negative values. In particular, we observe the reduction in the DQD
tunneling rate around zero detuning in response to the QPC noise,\cite{Aguado00} an effect which is associated with the
normalization of the spectral density of the dot wave functions.

\begin{figure}[b]
\includegraphics{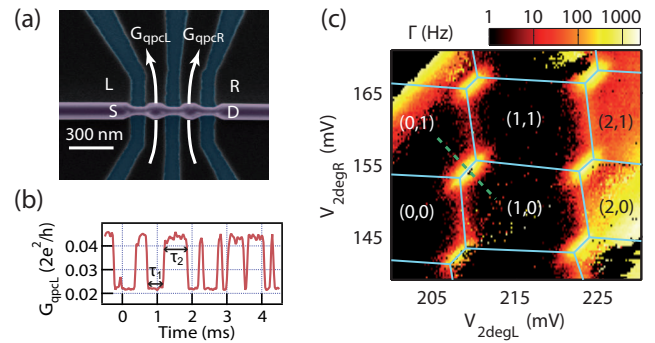}
\caption{(Color online) (a) SEM image of the device. A single etching step defines the DQD in the nanowire (horizontal)
and constrictions in the 2DEG underneath. The white arrows indicate the direction of negative current through the point
contacts. (b) Conductance $G_\mathrm{qpcL}$ of the left QPC measured time-resolved at a gate configuration where charge
exchange between the dots is energetically allowed. The QPC conductance drops whenever an electron tunnels from the
right into the left dot. (c) Charge stability diagram of the DQD obtained by evaluating the event rate $\Gamma =
1/\langle \tau_1+\tau_2 \rangle$ in traces as in (b). The solid lines delineate regions of stable occupation numbers
$(n,m)$ of the DQD relative to the arbitrary reference $(0,0)$. The subsequent measurements are carried out along the
dashed line. } \label{fig:NWPAT_figure_Intro}
\end{figure}

Figure \ref{fig:NWPAT_figure_Intro}(a) shows a scanning electron microscope (SEM) image of the sample. It is fabricated
by depositing an InAs nanowire on top of a GaAs/AlGaAs heterostructure containing a 37-$\mathrm{nm}$-deep
two-dimensional electron gas (2DEG; density $4 \times 10^{11} \, \mathrm{cm}^{-2}$, mobility $3 \times 10^5
\mathrm{cm^2/Vs}$ at $2 \, \mathrm{K}$). By subsequent electron beam lithography and wet etching, QDs in the nanowire
and constrictions in the 2DEG are defined simultaneously which ensures perfect alignment between QD and sensor (for
details see Refs.~\onlinecite{Shorubalko08} and \onlinecite{Choi08}). The parts of the 2DEG marked ``L'' and ``R'' serve as
side gates to tune the QPC conductances. Similarly, the QPCs are used as gates to selectively tune the QD potentials by
applying offset voltages $V_\mathrm{2degL}$, $V_\mathrm{2degR}$ to both source and drain. All measurements were done in
a $^4 \mathrm{He}$ cryostat at $T = 2 \, \mathrm{K}$.

The DQD is operated at zero source-drain voltage and is tuned to a regime with very opaque barriers where its charge
dynamics is monitored with the QPC sensors. These are biased with source-drain voltages $V_\mathrm{qpcL}$,
$V_\mathrm{qpcR}$, and their currents are measured with a bandwidth of $10 \, \mathrm{kHz}$. Figure
\ref{fig:NWPAT_figure_Intro}(b) shows a typical time dependence of the left QPC's conductance, which exhibits steps
whenever a dot-charging event takes place. In measuring the event rate $\Gamma = 1/\langle \tau_1+\tau_2 \rangle$ as a
function of $V_\mathrm{2degL}$ and $V_\mathrm{2degR}$, we expect to reproduce the DQD charge stability diagram with
nonzero $\Gamma$ along the boundaries of the hexagonal regions of stable charge. In the corresponding graph in
Fig.~\ref{fig:NWPAT_figure_Intro}(c), which contains the data from the left QPC readout, we observe nearly vertical
lines belonging to tunneling between the left QD and the source lead ($\sim 10 \, \mathrm{Hz}$) and short, diagonal
lines where interdot tunneling takes place ($\sim 1 \, \mathrm{kHz}$). The horizontal charging lines of the right dot
are only visible in the time-averaged signal (not shown), as the associated dot-drain transitions are too fast to be
resolved in real time. The absolute occupation numbers of the two dots are not known, and hence the index pairs $(n,m)$
assigned to the hexagons mark the excess electron numbers relative to the state labeled $(0,0)$.

Apart from the features associated with the DQD, counts are detected in the (2,0) and (2,1) regions, as well as in the
top left corner of the plot \ref{fig:NWPAT_figure_Intro}(c). We attribute these additional fluctuations to charge traps
residing in the vicinity of the QPCs. Performing the measurements in the crossover region between charge states (0,1)
and (1,0), we avoid disturbance by additional charge traps. We note that a change in QPC bias may trigger the activity
of additional fluctuators, which however would leave clear signatures in the time-resolved current traces. In
particular, additional amplitude- and time-scales characteristic for the dynamics of the charge trap would show up  in
the current traces on top of the DQD signal. In our experiments no such additional features have shown up.

\begin{figure}
\includegraphics{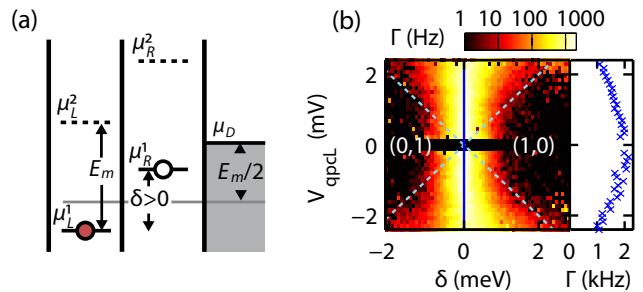}
\caption{(Color online) (a) Energy-level diagram of the DQD system and the drain lead. Upon adding a second electron to
the DQD, the levels $\mu_\mathrm{L}^1$, $\mu_\mathrm{R}^1$ shift by the mutual charging energy $E_m$ to the new positions $\mu_\mathrm{L}^2$,
$\mu_\mathrm{R}^2$. (b) Grayscale (colorscale) plot of the interdot tunneling rate $\Gamma$ as a function of level detuning
$\delta$ [dashed line in Fig.~\ref{fig:NWPAT_figure_Intro}(c) with $(\mu_\mathrm{L}^1+\mu_\mathrm{R}^1)/2=\mu_D-E_m/2$] and source-drain
voltage $V_\mathrm{qpcL}$ across the left QPC. Photons with energies bounded by $|e V_\mathrm{qpcL}|$ are emitted by
the QPC and can drive inelastic tunneling events which leads to a broadening of the main peak (dashed lines indicate
the condition $\delta = \pm e V_\mathrm{qpcL}$). The plot on the right is a cut through $\delta = 0\,\mathrm{meV}$
(solid line). } \label{fig:NWPAT_figure_SPhDet_Comic}
\end{figure}

In moving along the dashed line in Fig.~\ref{fig:NWPAT_figure_Intro}(c), we continuously vary the energy difference
$\delta = \mu_\mathrm{R}^1-\mu_\mathrm{L}^1$ between the charge configurations (1,0) and (0,1), as illustrated in the
level diagram in Fig.~\ref{fig:NWPAT_figure_SPhDet_Comic}(a). The energies $\mu_\mathrm{L}^2$ and $\mu_\mathrm{R}^2$
required for doubly occupying the DQD are higher by the mutual charging energy $E_m \approx 0.8 \, \mathrm{meV}$, which
was determined by finite-bias spectroscopy. An electron in the lower-energy dot can tunnel to the higher-energy dot by
absorbing an energy quantum $|\delta |$ from the environment. The DQD system therefore acts as a tunable and
frequency-selective probe for electrical noise in its vicinity.\cite{Aguado00}$^,$\cite{Gustavsson07a}

By applying a voltage $V_\mathrm{qpcL(R)}$ across one of the QPCs, we generate broadband noise with a high-frequency
cutoff at $eV_\mathrm{qpcL(R)}/h$, meaning that the electrons passing through the QPC have an exponentially small
probability to emit photons with energies higher than the bias energy.\cite{Khlus87, Lesovik89,
Zakka-Bajjani07}$^,$\cite{ Gustavsson07a} Due to the capacitive coupling, the generated photons can be absorbed by the
DQD and drive inelastic transitions. In measuring the interdot tunneling rate $\Gamma$ as a function of $\delta$ for
increasing QPC bias, we therefore expect the equilibrium tunneling peak at $\delta = 0$ to become broadened due to
photon absorption in a window $|\delta | < |e V_\mathrm{qpcL(R)}|$. In Fig.~\ref{fig:NWPAT_figure_SPhDet_Comic}(b), we
plot the corresponding measurement for bias applied across the left QPC. A small region $|V_\mathrm{qpcL}| < 0.15 \,
\mathrm{mV}$, where the signal-to-noise ratio of the counting signal is not sufficient, is excluded from the data.

A remarkable feature of the data in Fig.~\ref{fig:NWPAT_figure_SPhDet_Comic}(b) is the fact that not only the peak
\emph{width} is influenced by the QPC, but also its \emph{amplitude}. The maximum $\Gamma$ at $V_\mathrm{qpcL} = \pm 2
\, \mathrm{mV}$ is smaller by a factor of 0.6 compared to the maximum at $V_\mathrm{qpcL} \approx 0 \, \mathrm{mV}$.
Direct gating by the voltage $V_\mathrm{qpcL}$ can be excluded as the origin because of the magnitude of the effect,
and second because of the symmetry in positive and negative $V_\mathrm{qpcL}$. A similar reduction in the resonant
current through a QD as a function of QPC bias, has been reported in Ref.~\onlinecite{Onac06a}. There, the effect could
be explained by the excitation of an electron on the dot to a higher-energy state, from where it had the chance to
tunnel back to the source lead.

We propose that in our measurements dot-lead processes are not relevant; this will be justified in more detail later in
this paper. Instead, the reduction in $\Gamma$ at zero $\delta$ is directly linked to its increase at non-zero $\delta$
via shift of spectral weight. For the discussion of this effect, we consider the QPC as a source of voltage noise,
i.e., potential fluctuations $\hat{V}(t)$ across the central dot barrier with a spectral density which we denote
$S_V(\omega)$. Such fluctuations lead to inelastic tunneling through the central barrier, which is expressed in terms
of the probability density for the dot to exchange energy quanta $E$ with the source of the field,\cite{Devoret90}
\begin{equation}
\label{eq:NWPAT_stateSpectralDensity_Exponential} P(E) = \frac{1}{2\pi\hbar} \int \! \dd t \,
\mathrm{exp}\left[J(t)+\ii E t/\hbar \right].
\end{equation}
Here, the potential fluctuations are described in terms of the autocorrelation function $J(t)=\langle [\hat{\phi}(t) -
\hat{\phi}(0)] \hat{\phi}(0) \rangle$ of the phase operators $\hat{\phi}(t) = \int_0^t \dd t' e \hat{V}(t')/\hbar$.
Equation (\ref{eq:NWPAT_stateSpectralDensity_Exponential}) is valid for fields $\hat{V}(t)$ with typical frequencies
much larger than the tunneling rate through the barrier, a regime in which $P(E)$ can be interpreted as the spectral
density of the electronic state in either of the two dots. It is normalized to unity and determines the energy
dependence of the unidirectional tunneling rates between the dots, namely, $\Gamma_\mathrm{L \leftarrow R}(\delta)
\equiv \Gamma_\mathrm{LR}(\delta) \propto P(\delta)$ (right to left) and $\Gamma_\mathrm{RL}(\delta) \propto
P(-\delta)$ (left to right). Notably, the fact that $P(\delta)$ is not (necessarily) even in $\delta$ is due to the
voltage $\hat{V}(t)$ being a quantum mechanical operator which does not commute with itself at different times, and
hence $J(t) \neq J(-t)$ in Eq.~(\ref{eq:NWPAT_stateSpectralDensity_Exponential}). It is only within this (quantum)
formalism that $P(\delta)$ can describe spontaneous emission of energy quanta into the modes of the field; if the
voltage were classical, only stimulated absorption and emission would be possible and $P(\delta)$ would be even in
$\delta$.

\begin{figure}
\includegraphics{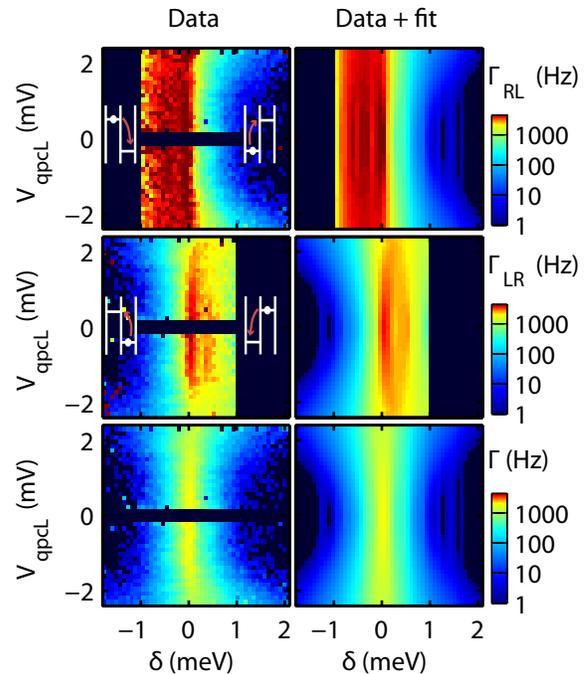}
\caption{(Color online) Left column: grayscale (colorscale) plots of the unidirectional tunneling rates
$\Gamma_\mathrm{LR,RL}$ and the total event rate $\Gamma$ as functions of detuning $\delta$ and QPC bias
$V_\mathrm{qpcL}$. Some of the data ($\Gamma_\mathrm{RL}$: $\delta<-1 \, \mathrm{meV}$; $\Gamma_\mathrm{LR}$: $\delta>1
\, \mathrm{meV}$) with large statistical errors have been removed. Right column: Theoretical
$V_\mathrm{qpcL}$-dependence of the data reconstructed by numerical convolution of the measured cuts at
$V_\mathrm{qpcL} = -460 \, \mathrm{\mu V}$ (ideally $V_\mathrm{qpcL} = 0 \, \mathrm{\mu V}$) with the spectral density
$P^\mathrm{ex}(E ;V_\mathrm{qpcL})$, cf.~Eq.~(\ref{eq:NWPAT_stateSpectralDensity}).}
\label{fig:NWPAT_figure_Convolution}
\end{figure}

We are interested in the inelastic tunneling processes caused by the biased QPC. Other inelastic processes, however,
are not negligible but instead even dominate over the QPC-driven processes. In particular, emission and absorption of
phonons\cite{Fujisawa98} within the InAs wire is known to be the most important mechanism for inelastic interdot
tunneling. Here, we treat these processes as a background noise present at zero QPC voltage. We therefore split the
voltage fluctuations $\hat{V}$, or rather their spectral density $S_V(\omega)$, into an \emph{equilibrium} part
$S_V^{(0)}(\omega)$, which includes thermal fluctuations in the nanowire as well as contributions from the QPC, and an
\emph{excess} part $S_V^\mathrm{ex}(\omega; V_\mathrm{qpcL})$, generated by the QPC at finite voltage
$V_\mathrm{qpcL}$.  The equilibrium contribution $S_V^{(0)}$ then determines the line shape of the rates
$\Gamma_\mathrm{RL,LR}$ at zero QPC voltage, while the finite-bias noise shifts weight to higher/lower energies. Since
the exponents $J(t)=J^{(0)}(t)+ J^\mathrm{ex} (t)$ for the equilibrium and the excess noise contributions are additive,
it follows from Eq.~(\ref{eq:NWPAT_stateSpectralDensity_Exponential}) that their probability densities $P^\mathrm{ex}(E
; V_\mathrm{qpcL})$ and $P^{(0)}(E)$ have to be convolved to obtain the total $P(E)$. Identifying $P(\delta;
V_\mathrm{qpcL}) \propto \Gamma_\mathrm{LR}(\delta ; V_\mathrm{qpcL})$ and $P^{(0)}(\lambda) \propto
\Gamma_\mathrm{LR}(\lambda ; 0)$ (and likewise for $\Gamma_\mathrm{RL}$), we can relate the rates at finite and zero
bias,
\begin{eqnarray}
\label{eq:NWPAT_newGammas_convolution}
  \Gamma_\mathrm{LR}(\delta ;V_\mathrm{qpcL}) = \int \dd \lambda \,
  \Gamma_\mathrm{LR}(\lambda ;0) P^\mathrm{ex} (\delta-\lambda ;V_\mathrm{qpcL}).
\end{eqnarray}
To leading order in $J^\mathrm{ex}$, the probability density $P^\mathrm{ex} (E;V_\mathrm{qpcL})$ can be
expressed\cite{Aguado00} through $S_V^\mathrm{ex}(\omega ; V_\mathrm{qpcL})$,
\begin{eqnarray}
\label{eq:NWPAT_stateSpectralDensity}
  P^\mathrm{ex}(E ;V_\mathrm{qpcL}) = \left[ 1-\frac{e^2}{\hbar^2} \int \! \dd \omega
  \frac{S_V^\mathrm{ex}(\omega ;V_\mathrm{qpcL})}{\omega^2}
   \right] \delta (E) \nonumber\\
  +\frac{e^2}{\hbar}\frac{S_V^\mathrm{ex}(E/\hbar ;V_\mathrm{qpcL})}{E^2}.
\end{eqnarray}
We assume $S^\mathrm{ex}_V$ to be proportional to the current shot-noise $S^\mathrm{ex}_I$ of the QPC, which amounts to
assuming a frequency-independent trans-impedance $Z_\mathrm{tr}$ relating the two. The symmetrized version of the
shot-noise has been calculated by Khlus\cite{Khlus87} and by Lesovik,\cite{Lesovik89} while the non-symmetrized
expression has been found in Refs.~\onlinecite{Lesovik97} and \onlinecite{Aguado00}. Depending on the specific detector
design, it is the non-symmetrized noise which is usually measured in an experiment, either at positive frequencies
only\cite{Lesovik97} or at both positive and negative frequencies, as is the case in Refs.~\onlinecite{Aguado00} and
\onlinecite{Bouchiat09} as well as here. In the latter case, the asymmetry between positive and negative frequency
results provides information on zero-point fluctuations (at sufficiently high frequencies as compared to the applied
voltage and temperature). In the present situation, the excess noise is symmetric in frequency and hence we do not get
access to those fluctuations (cf.\ Ref.\ \onlinecite{Bouchiat09} for a situation where a non-symmetric excess noise is
measured). The expression for the excess noise easily derives from the result in Ref.~\onlinecite{Aguado00} or, since
it is symmetric, from the original symmetrized results in Refs.~\onlinecite{Khlus87} and \onlinecite{Lesovik89},
\begin{eqnarray}
\label{eq:NWPAT_excessNoise}
  S_I^\mathrm{ex}(\omega;V_\mathrm{qpcL}) \!\!\! &= \frac{2e^2}{\hbar} D(1-D) \bigg[ -2\hbar \omega  \coth \left( \frac{\hbar \omega}{2k_BT}\right)  \nonumber\\
  &\:\:\:+\,(eV_\mathrm{qpcL}+\hbar \omega) \coth \left( \frac{eV_\mathrm{qpcL}+\hbar \omega}{2k_BT}\right) \quad \nonumber\\
  &\:\:\:+\,(eV_\mathrm{qpcL}-\hbar \omega) \coth \left( \frac{eV_\mathrm{qpcL}-\hbar \omega}{2k_BT}\right) \bigg],
\end{eqnarray}
where $D = G_\mathrm{qpcL}  h/(2 e^2)$ is the transmission coefficient of the
QPC. The noise spectrum (\ref{eq:NWPAT_excessNoise}) is even in $\omega$, with
a maximum at $\omega = 0$, and is characterized by a high-frequency cutoff
$|\omega| < |eV_\mathrm{qpcL}/\hbar|$ that is smeared by temperature.

Experimentally, we extract the rates $\Gamma_\mathrm{LR,RL}$ from traces as that shown in
Fig.~\ref{fig:NWPAT_figure_Intro}(b) by averaging the time the signal spends in the low- or high-current
state,\cite{Gustavsson06a} $\Gamma_\mathrm{RL} \approx 1/\langle \tau_1 \rangle$ and $\Gamma_\mathrm{LR} \approx
1/\langle \tau_2 \rangle$. In the left column of Fig.~\ref{fig:NWPAT_figure_Convolution}, we plot these rates as
functions of $V_\mathrm{qpcL}$ and $\delta$. [For clarity, we also include the plot of $\Gamma =
\Gamma_\mathrm{LR}\Gamma_\mathrm{RL}/(\Gamma_\mathrm{LR}+ \Gamma_\mathrm{RL})$ identical to that in
Fig.~\ref{fig:NWPAT_figure_SPhDet_Comic}(b).] Indeed they qualitatively exhibit the principal features expected from
Eq.~(\ref{eq:NWPAT_stateSpectralDensity}) combined with the spectrum (\ref{eq:NWPAT_excessNoise}), namely, the
reduction in their maxima around $\delta = 0$ and their increase on the excitation side of the $\delta$ axis (that is,
$\delta > 0$ for $\Gamma_\mathrm{RL}$, and $\delta < 0$ for $\Gamma_\mathrm{LR}$).

For the quantitative comparison between experiment and theory, we simulated the effect of the QPC by numerically
performing the convolution of the energy density (\ref{eq:NWPAT_stateSpectralDensity}) with the measured rates at a QPC
bias close to zero, according to Eq.~(\ref{eq:NWPAT_newGammas_convolution}). The only unknown parameter in this
analysis is the trans-impedance $Z_\mathrm{tr}$ in $S_V^\mathrm{ex}(\omega;V_\mathrm{qpcL}) =
|Z_\mathrm{tr}|^2S_I^\mathrm{ex}(\omega ;V_\mathrm{qpcL})$ that was determined by minimizing the fitting error
(weighted according to the inverse experimental uncertainty). Some care has to be taken concerning the coefficient $D$
appearing in the noise spectrum (\ref{eq:NWPAT_excessNoise}). As seen in Fig.~\ref{fig:NWPAT_figure_Intro}(b), the
relative changes in $G_\mathrm{qpcL}$ caused by the hopping dot electron are large and the $D$ coefficient relevant for
the L$\rightarrow$R processes (corresponding to the low-current state of the QPC signal, $D_\mathrm{L} = 0.021$) is
therefore significantly different from the one relevant for R$\rightarrow$L processes ($D_\mathrm{R} = 0.045$). The
result of the analysis is given in the bottom row of Fig.~\ref{fig:NWPAT_figure_Convolution} and shows a good agreement
between theory and experiment. Both data sets, $\Gamma_\mathrm{LR}(\delta ; V_\mathrm{qpcL})$ and
$\Gamma_\mathrm{RL}(\delta ; V_\mathrm{qpcL})$, are best approximated using a trans-impedance of $Z_\mathrm{tr} = 5.4
\, \mathrm{k \Omega}$. This value is roughly one order of magnitude larger than the corresponding figure given in
Ref.~\onlinecite{Onac06a}, a fact which is well explained with the stronger capacitive coupling in the present case.

In order to rule out alternative explanations of the data, we shortly discuss how QPC-driven tunneling between right
dot and lead may affect the measurement of $\Gamma_\mathrm{LR,RL}$, and may account for the reduction in the tunneling
rates near $\delta = 0 \, \mathrm{meV}$. Namely, when allowing for such dot-lead tunneling, the rate
$\Gamma_\mathrm{LR}(\delta \approx 0 \, \mathrm{meV})$ would become smaller for increasing bias $V_\mathrm{qpcL}$
because an electron on the right dot may be excited into the lead (and tunnel back) instead of tunneling into the left
dot. In the case of $\Gamma_\mathrm{RL}(\delta \approx 0 \, \mathrm{meV})$ instead, the possible mechanism would
involve excitation \emph{into} the dot: the QPC may overcome the mutual charging energy and excite an electron from the
lead into the right dot, thus blocking tunneling from left to right dot [see also
Fig.~\ref{fig:NWPAT_figure_SPhDet_Comic}(a)].

However, such dot-lead processes take place predominantly as compositions of excitations to an intermediate state and
subsequent elastic tunneling into the lead\cite{Gustavsson08b} (or, in the other direction, excitation of the valence
electron of the dot and subsequent tunneling from the lead into the unoccupied low-energy state). This is in
contradiction to the fact that there is no indication of an onset in bias voltage in the data, which should be present
at the energy of the excited state, if dot-lead processes were relevant. Instead, the changes in
$\Gamma_\mathrm{RL,LR}$ are smooth and gradual, and already significant at $eV_\mathrm{qpcL}$ values below the typical
single-particle excitation energy of $0.8 \, \mathrm{meV}$ in our sample. Furthermore, we stress that the PAT theory is
able to quantify the behavior of the rates on and off peak with a \emph{single} parameter, whereas in a model
incorporating dot-lead processes these two are separate regimes.

\begin{figure}
\includegraphics{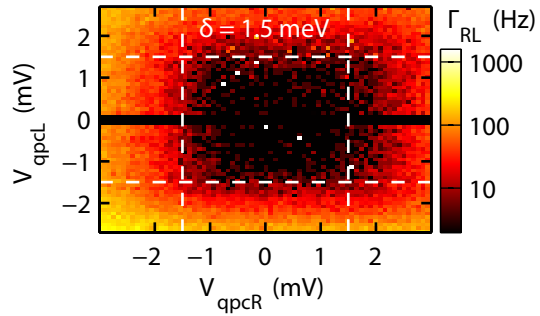}
\caption{(Color online) Photon-absorption rate $\Gamma_\mathrm{RL}$ as a function of the left and right QPC
source-drain voltages measured at fixed detuning $\delta = 1.5 \, \mathrm{meV}$ (dashed lines).}
\label{fig:NWPAT_figure_GammaVsBothQPCs}
\end{figure}

For all measurements presented up to now, the right QPC's source-drain voltage was set to zero. Similar to the left
QPC, it is tuned to a low conductance of approximately $0.08e^2/h$, and can be used as a noise source. This allows to
test whether the perturbations created by the two QPCs have independent effects on the dot. To this end, we fix the
detuning at a value of $\delta = 1.5 \, \mathrm{meV}$ and measure $\Gamma_\mathrm{RL}$ (i.e., the photon-absorption
rate) as a function of left and right QPC bias voltages. The result, shown in
Fig.~\ref{fig:NWPAT_figure_GammaVsBothQPCs}, is a plot with a characteristic, square structure with a region of zero
absorption in the inner part. This confirms the picture described above in the sense that there is a well-defined
energy threshold for one-photon processes $(|eV_\mathrm{qpcL}|,|eV_\mathrm{qpcR}|)>|\delta|$, and that to first order
the effects of the two QPCs add independently. Two-photon processes with photons arriving at the double dot originating
from the two QPCs would result in an additional, possibly diamond-like structure due to the differing energy
conservation condition $|e V_\mathrm{qpcL}| + |e V_\mathrm{qpcR}| \geq |\delta|$. A measurement of the kind shown in
Fig.~\ref{fig:NWPAT_figure_GammaVsBothQPCs} provides a unique way of directly mapping the addition of energies of
photons\cite{Tobiska06} emitted by different sources.

We have investigated the process of photon-assisted tunneling driven by QPC noise in an InAs based DQD. Due to the full
tunability of the DQD, we could observe the expected suppression of tunneling for zero dot detuning with increasing
noise strength compensating for the increase in tunneling for nonzero detuning. Our data can be understood by treating
the QPC as a high-frequency noise source. Finally, by measurements with two separate emitter QPCs we confirm that their
effects add up independently.

The authors thank F.~Portier, G.~Lesovik, and S.~Ludwig for fruitful discussion. Financial support from the Swiss
National Science Foundation (Schweizerischer Nationalfonds) is gratefully acknowledged.

\bibliographystyle{apsrev}

\end{document}